\documentclass[twoside,11pt]{article}

\usepackage{jmlr2e}
\usepackage{hyperref}
\usepackage{amsmath,amsfonts}
\usepackage{bbm}
\usepackage{multirow}
\usepackage{soul}
\usepackage{enumitem}

\jmlrheading{1}{2000}{1-48}{4/00}{10/00}{}

\firstpageno{1}

\begin{document}

\title{Disentangled Representation for Diversified Recommendations}

\author{\name Xiaoying Zhang \email zhangxiaoying.xy@bytedance.com \\
       \addr AI Lab, Bytedance Inc. 
       \AND
       \name Hongning Wang \email hw5x@virginia.edu \\
       \addr Department of Computer Science \\ University of Virginia, USA
       \AND
       \name Hang Li \email lihang.lh@bytedance.com \\
       \addr AI Lab, Bytedance Inc.
       }

\editor{}
\maketitle

\begin{abstract}
%Accuracy and diversity are long believed two conflicting goals for recommendations.
Accuracy and diversity have long been considered to be two conflicting goals for recommendations. 
We point out,  however,  that as the diversity is typically measured by certain pre-selected item attributes, e.g., category as the most popularly employed one, improved diversity can be achieved without sacrificing recommendation accuracy, as long as the diversification respects the user's preference about the pre-selected attributes.
This calls for a fine-grained understanding of a user's preferences over items, where one needs to recognize the user's choice is driven by the quality of the item itself, or the pre-selected attributes of the item. 

In this work, we focus on diversity defined on item categories. We propose a general diversification framework agnostic to the choice of recommendation algorithms. 
Our solution
%employs an adversarial discriminator to
disentangles the learnt user representation in the recommendation module into category-independent and category-dependent components to differentiate a user's preference over items from two orthogonal perspectives. 
Experimental results on three benchmark datasets and online A/B test demonstrate the effectiveness of our solution in improving both recommendation accuracy and diversity.
In-depth analysis suggests that the improvement is due to our improved modeling of users' categorical preferences and refined ranking within item categories.

\end{abstract}


\section{Introduction}
Recommender systems learn users' interests from historical observations (e.g., their clicks, bookmarked or purchased items, etc.) so as to identify the items that best suit users' preferences.
The success of recommender system in enhancing user experience and boosting platform utility has been witnessed in a number of scenarios including
e-commerce~\citep{zhou2018deep,he2017neural}, online news recommendation~\citep{wu2019npa} and streaming services~\citep{covington2016deep}.

Recommendation accuracy, which measures whether a recommendation model can recommend items that  users will like, serves as the dominant
target or even the only target in most previous work~\citep{zhou2018deep,wang2021deconfounded,he2017neural,guo2017deepfm,covington2016deep}.
Various complicated models~\citep{zhou2018deep,guo2017deepfm,covington2016deep} have been proposed for higher accuracy.
While recommendation accuracy has been shown to be closely related to user satisfaction, 
it is never the only rule of thumb.
Recent work found the recommendation diversity,
which measures the dissimilarity among recommended items regarding certain pre-selected item attributes (e.g., item category)
%with category as the most popularly employed criterion, 
%which measures dissimilarity of recommended items, 
also plays an important role in the overall user experience~\citep{wilhelm2018practical,kapoor2015like,zhou2010solving}.
For example, even if a user is a fan of  basketball,
he/she can still get bored with recommendations only about basketball videos or news,
which increases the risk of user attrition.

\begin{figure}[t]
    \centering  \includegraphics[width=0.8\linewidth]{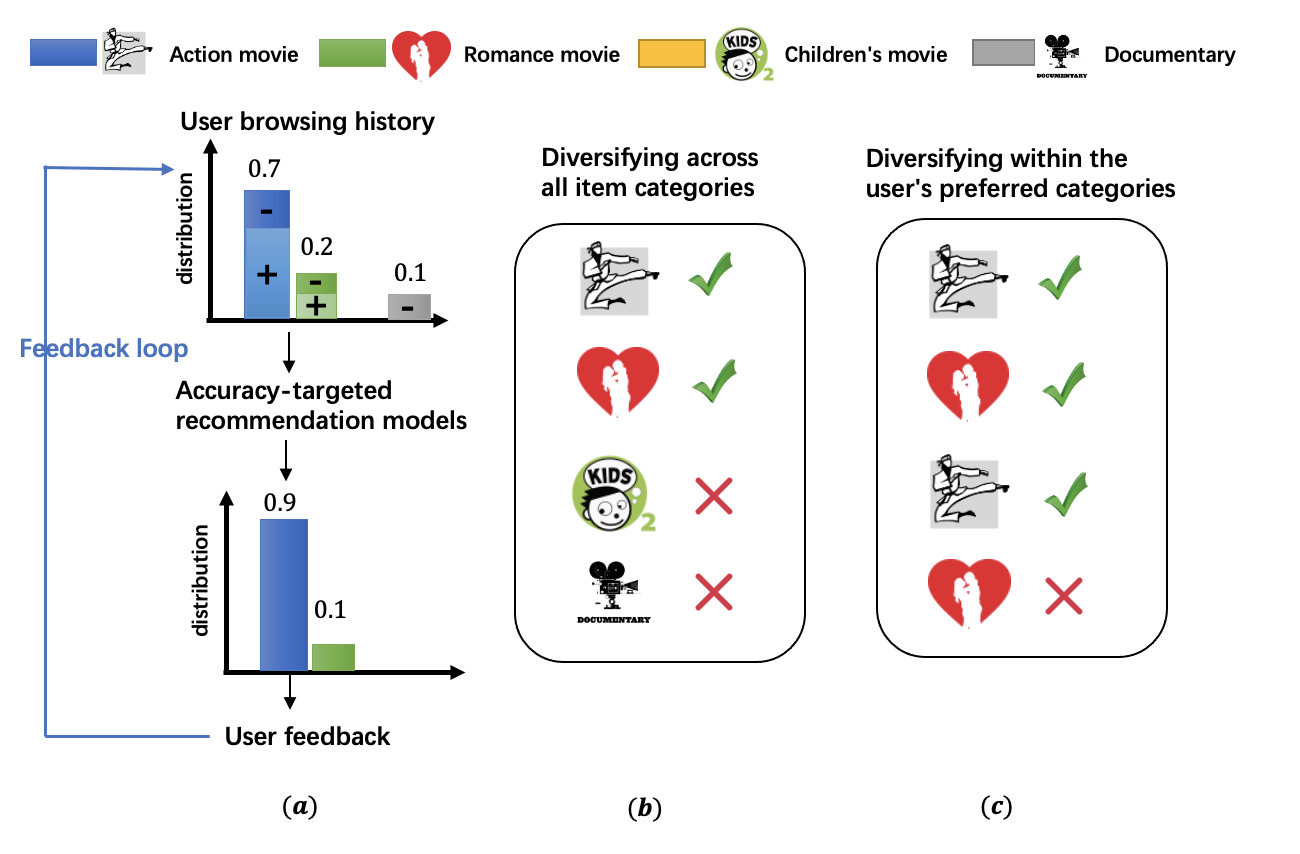}
    \caption{Illustration of recommendation accuracy and diversity optimization in different recommendation models.}
    \label{fig:example}
\end{figure}

Following previous work~\citep{steck2018calibrated,wang2021deconfounded,zheng2021dgcn}, we focus on diversity defined on item categories in this paper and aim to address the so-called accuracy-diversity dilemma~\citep{zheng2021dgcn}.
On one hand,  recommendation models with accuracy as their primary target often lose diversity to some extent, due to  overly emphasizing items in the dominant categories in a user's interaction history~\citep{steck2018calibrated,wang2021deconfounded}.
Figure~\ref{fig:example}(a) illustrates this issue with an example in movie
recommendation,
where 70\% of the movies watched by a user are
action movies, which leads 90\% of the system's recommendations to fall in the action movie category.
Worse still, because of the feedback loop~\citep{chaney2018algorithmic}, the emphasis on the dominant categories in the system's recommendations will be further intensified when the user follows the recommendations, causing further decreased recommendation diversity and issues like filter
bubbles~\citep{nguyen2014exploring} and echo chambers~\citep{ge2020understanding}.
On the other hand, simply diversifying recommendations over all item categories without considering the user's categorical preference hurts the accuracy of generated recommendations~\citep{wilhelm2018practical,ziegler2005improving,qin2013promoting,zheng2021dgcn}.
As shown in Figure~\ref{fig:example}(b), although the recommendation list is diverse by covering all four categories, negative feedback is more likely on the categories where the user interacted less often or negative feedback already prevailed, e.g., children's movies and documentary movies respectively in this example.
%One question is whether there is a novel recommendation model that can contribute to both accurate and diversified recommendation?

%\hnote{Can we create an illustrative figure to demonstrate the necessity/benefit in disentangling the representation?}

%As recommendation diversity is often measured by pre-selected item attributes, i.e., item category, one can improve recommendation diversity without hurting the accuracy of generated recommendations, as long as diversification is conducted within item categories that the user may prefer.
Clearly one should not  recklessly increase diversity. For categories the user is less likely to be interested in, the risk of making a bad recommendation overweights the benefit of increased diversity.
Thus, this paper focuses on conducting diversification only among item categories that the user prefers, suggesting the possibility to improve recommendation diversity without sacrificing recommendation accuracy.
% This also suggests the possibility to improve recommendation diversity without sacrificing recommendation accuracy, as long as diversification is conducted among the item categories that the user prefers.
Figure~\ref{fig:example}(c) gives an example recommendation list following such a strategy, where the recommended items mainly fall in action and romance movies, the two preferred categories inferred from the user's interaction history.
%of the user.
This strategy requires the recommendation model to clearly distinguish whether the user's positive/negative feedback is due to the item's category or other category-independent
features of the item (e.g., the item's own quality), which was ignored by previous recommendation models.
%However, these two levels of preference modeling are mixed up in previous recommendation models.
% We observe that 
% In other words, given a user's feedback to an item,  previous recommendation models cannot measure how much effect the user's preference over the item's category and over act on the feedback respectively.
% Such entanglement decreases both the diversity and accuracy of subsequent recommendations.
% For example, with such entanglement, previous recommendation models may  misinterpret the popularity of dominant category as users' high preference over the category,
% leading to over-emphasize items in the dominant category,
% and even give high prediction scores to items of low-interest or low-quality in the dominant category.
% For another example, under such entanglement, if a user interacts with several bad items in a category, 
% then those negative feedbacks may also be taken as the user's low preference over the category,
% decreasing the probability to further recommend good items in this category.

In this paper, we propose a general and model-agnostic framework to \underline{d}isentangle a user's \underline{c}ategory-dependent and category-independent preferences for an accurate and diversified \underline{r}ecommender \underline{s}ystem (DCRS).
Specifically, DCRS takes a user's preference over an item as a product of:
(1) the user's preference over the item's category; and
(2) the user's preference over category-independent features of the item,  e.g., the item's quality.
Such disentanglement suggests a hierarchical decision making process by the user:
If a user has a strong preference over a particular category of items, he/she may still enjoy items of this category, even though their qualities are not perfect.
However, if the probability that a user likes a category is low,
%if a user has a low preference in a category,
only items of high quality in this category could have a chance to be considered.
The disentanglement  ensures items of the same quality, but in different categories that a user prefers similarly, have equal probabilities to be recommended. It naturally avoids overly recommending items from the dominant categories  in the  user's interaction history.
The main challenge therefore lies in how to disentangle a user's  preference regarding the aforementioned two orthogonal perspectives, given
his/her preference over the item categories is not observable.
This makes naive solutions like using different supervision signals to separately train users' representations \citep{zheng2021disentangling},
or separating items' feature vectors into  category dependent and independent segments, ineffective.

%Moreover, since item category serves as a key item feature as shown in previous work~\citep{kang2021learning}, directly removing category-dependent features will greatly hurt the accuracy of recommendation.

%Inspired by recent advances in adversarial learning~\citep{bose2019compositional,li2021towards}, 
DCRS is agnostic to the choice of recommendation module, which is supposed to learn informative representations of users and items.
In particular, DCRS adopts a discriminator to disentangle the learnt representation into  category-independent and category-dependent segments respectively.
The recommendation module and discriminator are learnt simultaneously to ensure the effectiveness of disentangled representation learning for accurate and diverse recommendations.
%{\color{blue}
To evaluate the proposed DCRS solution, we conduct both offline experiments on three benchmark datasets and online A/B test on 
Toutiao app,
one of  the largest news recommendation platforms in China.
%\footnote{The name of the platform is temporarily anonymized due to double-blind review policy.}.
Experiment results demonstrate that DCRS can successfully recommend diverse items that users prefer, and thus improve both recommendation accuracy and diversity. 
In-depth analysis and case studies suggest strong evidence showing:
(1) the disentangled category-independent representation from DCRS can distinguish the user's preference within category more accurately; and 
(2) DCRS can capture a user's diverse preferences in historical interactions more thoroughly.
All codes and data can be found in 
\url{https://github.com/Xiaoyinggit/DCRS.git}.
% supplementary materials \citep{dcrs-supple} for reproducibility.
%to ensure the  of this work.
%}

Overall, our contribution of this work is as follows:
\begin{itemize}
    \item We demonstrate that accuracy and diversity are not conflicting goals for recommendation, as long as the diversification respects the user's categorical preference.
    \item To capture a user's latent preferences on item categories more accurately, our proposed DCRS disentangles the user's preference into category-dependent and category-independent components.
    \item Experiments on three benchmark datasets and online A/B test demonstrate the effectiveness of DCRS in improving both recommendation accuracy and diversity. In-depth analysis further demonstrates the improvement comes from more accurate modeling of the user's preference both over and within categories. 
\end{itemize}

\section{Framework}
In this section, we describe how the proposed DCRS solution disentangles a user's category dependent and independent preferences to simultaneously improve recommendation accuracy and diversity. 
For the ease of illustration, we first briefly describe a general architecture which covers almost all popularly used recommendation models.
We then depict how to smoothly integrate DCRS into such a general architecture to diversify its recommendations.

\subsection{Preliminary: A General Recommendation Architecture}
\label{subsec:preliminary}
In a recommendation task, we are given a user behavior dataset $\boldsymbol{\mathcal{X}}$ that contains interactions between $N$ users and $M$ items. 
The interaction between user $u$ and item $i$ is represented as a tuple $(u,i, y_{u,i}) \in \boldsymbol{\mathcal{X}}$.
Here $y_{u,i} \in \{0,1\}$  denotes user $u$'s feedback to item $i$, where $y_{u,i}=1$ denotes  positive feedback (e,g., a click or a positive rating), and $y_{u,i}=0$ denotes negative feedback.
Generally speaking,  a recommendation model will first learn a user-item representation to capture the user's preference over the item:
\begin{equation}
\boldsymbol{h}_{u,i} = f(u,i,\boldsymbol{\theta}) \in R^d,
\label{equ:h_ui}
\end{equation}
where $\boldsymbol{\theta}$ denotes a set of trainable parameters in the recommendation model.
Various architectures~\citep{zhou2018deep, wang2021deconfounded, he2017neural,guo2017deepfm} have been proposed to implement  $f(u,i,\boldsymbol{\theta})$, ranging from the simple matrix factorization algorithm~\citep{mnih2007probabilistic} that directly takes the element-wise product of user and item  embeddings to form the representation,  to complex architectures such as the bi-interaction layer in NFM~\citep{he2017neural}.
Let $\hat{p}_{u,i}$ denote the probability that user $u$ gives  positive feedback to item $i$.
The goal of the recommendation model is to use the learnt user-item representation to estimate $\hat{p}_{u,i}$, either by directly summing up elements in $\boldsymbol{h}_{u,i}$ as in matrix factorization, or through a learnable projection layer as follows:
\begin{equation}
  \hat{p}_{u,i} = P \left(Y_{u,i}=1 |u,i\right)  = \sigma \left( \boldsymbol{W}^\top\boldsymbol{h}_{u,i} \right),
  \label{equ:pui}
\end{equation}
where 
$Y_{u,i}$ is a random variable representing the feedback from user $u$ on item $i$;
$\boldsymbol{W} \in R^{d\times 1}$ is the learnable weight vector of the projection layer, and $\sigma(\cdot)$ is the sigmoid function.
The parameters of the recommendation model are then optimized by minimizing the following loss:
\begin{equation}
\label{eq-obj}
    \mathcal{L}(\boldsymbol{\mathcal{X}}, \boldsymbol{\theta}, \boldsymbol{W}) =  \frac{1}{|\boldsymbol{X}|} \sum_{(u, i, y_{u,i}) \in \boldsymbol{\mathcal{X}}} \mathcal{L}_{rec}(y_{u,i}, \hat{p}_{u,i}), 
\end{equation}
where $\mathcal{L}_{rec}(\cdot, \cdot)$ represents the chosen loss function. 
Various loss functions have been explored in literature, inlcuding cross entropy loss, Mean Squared Error (MSE) and BPR loss~\citep{rendle2012bpr}. In this work,
%without further specification,
we will use the cross entropy loss by default.

\subsection{Disentangle Category Dependent and Independent Representations}
\label{sub_sec:dis_cat}
%\hnote{We need a high-level illustrative figure to explain our insight, preferably a hierarchical model to illustrate the generative process specified by (5a)-(5c).}

\begin{figure}[t]
    \centering\includegraphics[width=0.8\linewidth]{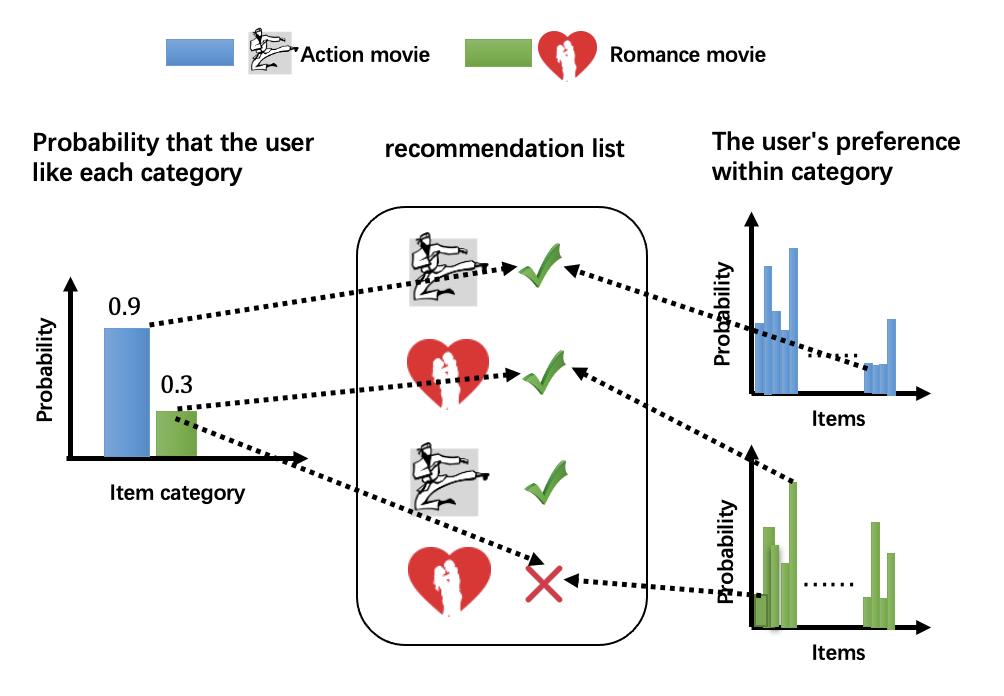}
    \caption{Hierarchical decision making process of DCRS framework. Each feedback is determined by: (1) the user's preference over the item's category; and (2) the user's preference over category-independent features of the item.  }
    \label{fig:framework}
\end{figure}

We consider a user's feedback on an item as a mixture reflecting his/her preference over the item's category and category-independent properties, e.g., the item's intrinsic quality.
As shown in Figure~\ref{fig:framework}, the first action movie that receives positive feedback can very likely be caused by the user's strong preference over the category of action movies, while his/her positive feedback on the second romance movie is more likely to be caused by its high quality that makes up the low probability that the user likes romance movies.
In order to diversify the recommendations with respect to a user's preferred categories, the recommendation model needs to clearly distinguish the effect of item category and other category-independent properties on a user's decision making. %, based on every piece of user's specific feedback.
To make our method description general enough to cover situations where an item can associate with multiple categories, we take item $i$'s category as the set that contains all categories that the item relates to, and denote it as $t_i$.
For example, assume there are three categories $\{c_1, c_2, c_3\}$ in a dataset.
If item $i$ is related to the first category, then $t_i=\{c_1\}$.
And if item $i$ is associated with the first two categories, then  $t_i=\{c_1, c_2\}$.
% For simplicity of notation, we take a $K$-dimensional vector $t_{i} \in R^{K}$ to represent item $i$'s category over $K$ number of unique  categories.
% For example, , then $t_{i}=[1,0,0]^\top$. 
% And if item $i$ is related to the first two categories, then $t_{i}=[0.5, 0.5, 0]^\top$.

% Previous recommendation models  entangle users' preference over categories and over category-independent features, and cannot distinguish how much effect the user's preference over category contribute to each feedback, as well as how much effect the user's preference over other category-independent features of the item act on each feedback.
% Thus they may mistake the popularity of dominant category in users' interaction history as users’
% high preference over the dominant category, leading to overrecommend items in the dominant category~\citep{steck2018calibrated,wang2021deconfounded}.
% Meanwhile negative feedbacks on bad items in a category may also wrongly interpreted as users' low reference over the category, discourging  further recommendation in the category.

We propose to disentangle  a user's preference over an item into two parts :
\begin{itemize}
    \item \textbf{Category-dependent preference}: it captures the user's preference over the item's category;
    \item \textbf{Category-independent preference}: it depicts how category-independent features affect the user's preference about the item.
    %, which determines the user's preference over items within category.
\end{itemize}
Such a disentanglement can be explained through a probabilistic view about the generation of user $u$'s feedback on item $i$.
Let $Y_{u,i}^C$  denote the binary random variable indicating user $u$'s feedback on item $i$'s category. We have the following,
\begin{subequations}
\begin{align*}
\!\!\!    P(Y_{u,i} =1| u, i) &= P(Y_{u,i}=1,  Y_{u,i}^C=1 | u, i)  \tag{4a} \\
    & =  P(Y_{u,i}=1 | u, i, Y_{u,i}^C=1)  P(Y_{u,i}^C=1| u,i)  \tag{4b}\\
%    & =  P(Y_{u,i}=1 | u, i, Y_{u,i,c}=1) \sum_{t_{i,c}} P(Y_{u,i,c}=1 | u,i, t_{i,c}) P(t_{i,c} | u,i)  \tag{5b}\\
%    & =  P(Y_{u,i}=1 | u, i, Y_{u,i,c}=1) \sum_{t_{i,c}} P(Y_{u,i,c}=1 | u, t_{i,c}) P(t_{i,c} |i)  \tag{5b}\\
    & = P(Y_{u,i}=1 | u, i, Y_{u,i}^C=1) P(Y_{u,i}^C=1|u, t_{i})  \tag{4c}
\label{equ:P}
\end{align*}
\end{subequations}
In particular, Eq.(4a) is due to the assumption that user $u$ gives positive feedback to item $i$ only if user $u$ likes item $i$'s category, i.e., $P(Y_{u,i}=1,  Y_{u,i}^C=1 | u, i)=1$ and $P(Y_{u,i}=1,  Y_{u,i}^C=0 | u, i)=0$.
Eq.(4b) follows the chain rule.
And Eq.(4c) is because $Y_{u,i}^C$ only depends on the item's category, instead of specific items. 
%$t_{i,c}$ only depends on item $i$  and $P(t_{i,c} | i)=1$.

The first term in  Eq.(\ref{equ:P}) depicts how likely user $u$ will give positive feedback to item $i$ when he/she is interested in item $i$'s category; and the second term models how likely user $u$ is interested in item $i$'s category.
Given that user $u$ likes the category of item $i$,  the probability in the first term only depends on the category-independent features of item $i$, such as item $i$'s quality, price, etc.
Thus, under the disentangled user-item representations, we can compute the first term by the probability $P(Y_{u,i}^{\perp C} =1 | u,i)$, which depicts user $u$'s preference over item $i$ driven by the \emph{category-independent features}. 
Thus, Eq.(\ref{equ:P}) can be rewritten as:
\begin{equation}
  \textstyle P(Y_{u,i}=1 | u, i) = P(Y_{u,i}^{\perp C} =1 | u,i)P(Y_{u,i}^C=1|u, t_{i}).
   \label{equ:P_distenglemnt}
\end{equation}
Eq.(\ref{equ:P_distenglemnt}) depicts a hierarchical decision making process illustrated in Figure~\ref{fig:framework}.
If user $u$ likes item $i$'s category with a higher probability $P(Y_{u,i}^C=1 | u, t_{i})$, he/she may still enjoy item $i$ even though item $i$'s quality is not perfect, indicated by a lower $P(Y_{u,i}^{\perp C} =1 | u,i)$.
For example, the positive feedback of the first action movie in Figure~\ref{fig:framework} is
%can be considered as
generated under such a scenario. 
Meanwhile, if there is only a small probability that user $u$ would be interested in item $i$'s category (i.e., low  $P(Y_{u,i}^C=1 | u, t_{i})$), item $i$ must be of high quality to get positive feedback, i.e.,  high  $P(Y_{u,i}^{\perp C} =1 | u,i)$. The positive feedback on the second romance movie in Figure~\ref{fig:framework} is a good example of this case.

% Why Eq.(5) can diversified within users' preferred categories.
Eq.(\ref{equ:P_distenglemnt}) also suggests why disentanglement makes recommendations diversified within a user's preferred categories.
Assume there are two categories $c_1$ and $c_2$ on which the user has similar preference.
Instead of recommending more items from the dominant category (either $c_1$ or $c_2$), 
via the disentanglement in Eq.(\ref{equ:P_distenglemnt}), 
items of the same quality within $c_1$ and $c_2$  will have an equal chance to be recommended, thus diversifying the recommendations.

% %, and items of category $c_1$ receives more positive interactions in training data.
% Without disentanglement, more positive interactions will be interpreted as higher preference on category $c_1$, thus recommending more items of category $c_1$.

Unfortunately, both terms in Eq.(\ref{equ:P_distenglemnt}) cannot be learnt via direct supervision signals, 
since neither user $u$'s feedback on item $i$'s category nor feedback driven by category-independent features of item $i$ can be observed.
% since user $u$'s feedback driven  solely  by item $i$'s category or its category-independent features cannot be observed.
Classical solutions would appeal to Expectation Maximization type algorithms~\citep{dempster1977maximum} to estimate the two terms in an iterative manner. However, given modern recommendation algorithms are usually realized via complex deep neural networks, posterior inference becomes cumbersome and also leads to slow convergence. Instead, DCRS implements Eq.(\ref{equ:P_distenglemnt}) by simultaneously learning two disentangled representations for estimating the two terms separately.
Specifically, DCRS learns two disentangled representations by:
%as follows,
\begin{equation}
\textstyle  \left[\left(\boldsymbol{h}_{u,i}^{\perp C}\right)^\top, \left(\boldsymbol{h}_{u,i}^C \right)^\top \right]^\top  = f(u,i,\boldsymbol{\theta}) \in R^{2d},
   \label{equ:disentangle_hui}
\end{equation}
where $\boldsymbol{h}_{u,i}^{\perp C} \in R^d$ aims to capture user $u$'s preference over category-independent features  to estimate $P(Y_{u,i}^{\perp C} =1 | u,i)$, 
and $\boldsymbol{h}_{u,i}^C \in R^d$ depicts user $u$'s preference over item $i$'s category $t_{i}$, aiming to estimate $P(Y_{u,i}^C=1|u, t_{i})$.
%One challenge is that how to force $\boldsymbol{h}_{u,i, \perp c}$ and $\boldsymbol{h}_{u,i, c}$ to focus on category-independent and  category-dependent features respectively.

Simply splitting item $i$'s feature vector into two parts, even with separate networks, cannot ensure complete disentanglement.
Instead, in addition to requiring the learnt the representations to best capture the user's preference,
we employ an adversarial discriminator that enforces  the learnt $\boldsymbol{h}_{u,i}^{\perp C}$ and $\boldsymbol{h}_{u,i}^C$ to be category-independent and category-dependent  respectively.
% The recommendation module and adversarial discriminator are learned simultaneously during training.

\begin{figure}[t]
    \centering
  \includegraphics[width=0.8\linewidth]{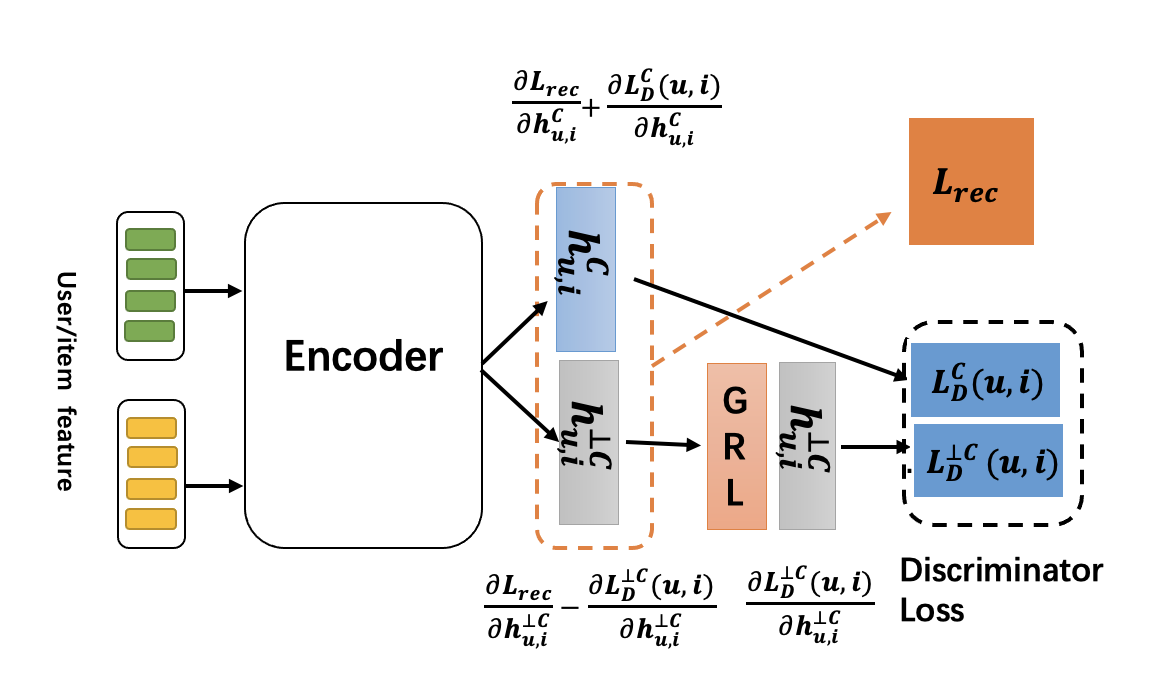}
    \caption{The architecture of DCRS, which disentangles the user $u$'s preference on item $i$ into category-dependent segment $\boldsymbol{h}^C_{u, i}$ and category-independent segment $\boldsymbol{h}^{\perp C}_{u,i}$ for diverse and accurate recommendations. }
    \label{fig:dcrs}
\end{figure}

 \noindent
 \textbf{Discriminator Module.} The discriminator $D(\cdot)$ acts as a category classifier, which takes one segment of disentangled representation, such as $\boldsymbol{h}_{u,i}^C$ or $\boldsymbol{h}_{u,i}^{\perp C}$, as input, and aims to predict the category of item $i$ (i.e., $t_i$).
 However, it is hard for the discriminator to directly predict $t_i$, since $t_i$ can take $2^K$-1 values, where $K$ is the number of unique categories available in the dataset. 
 For ease of learning, we represent $t_i$ by a  vector over $K$ unique categories, denoted as $\tilde{\boldsymbol{t}}_i$. 
 Again, assume there are three categories $\{c_1, c_2, c_3\}$, if $t_1=\{c_1\}$, then $\tilde{\boldsymbol{t}}_i=[1,0,0]^\top$.
 And if $t_1=\{c_1, c_2\}$, then $\tilde{\boldsymbol{t}}_i=[0.5,0.5,0]^\top$.
 Specifically, when relevance between item $i$ and each associated category can be measured \citep{pu2020multimodal}, a more accurate $\tilde{\boldsymbol{t}}_i$ can be achieved by making the $j$-th element of $\tilde{\boldsymbol{t}}_i$ proportional to the relevance between item $i$ and the $j$-th category.
 Otherwise, $\tilde{\boldsymbol{t}}_i$ can be simply assumed to be evenly distributed among related categories, which is also the default setting in our experiments. The discriminator then takes $\boldsymbol{h}_{u,i}^C$ or $\boldsymbol{h}_{u,i}^{\perp C}$ as input to predict $\tilde{\boldsymbol{t}}_i$.
 In our experiments, the discriminator $D(\cdot)$ is implemented via a fully connected layer, and  
  %and use cross entropy loss for its parameter estimation.
it should enforce the following: 
 \begin{itemize}
 \item Given $\boldsymbol{h}_{u,i}^C$ is closely related to item $i$'s category, the discriminator should predict $\tilde{\boldsymbol{t}}_i$  accurately based on   $\boldsymbol{h}_{u,i}^C$, i.e., the following loss should be minimized: 
  \begin{equation}
      \min \mathcal{L}_D^{C}(u,i) = \mathcal{L}_{CE} \left( D(\boldsymbol{h}_{u,i}^C),  \tilde{\boldsymbol{t}}_i\right), \nonumber
  \end{equation}
  where $\mathcal{L}_{CE}$ is the cross entropy loss.
  \item Given $\boldsymbol{h}_{u,i}^{\perp C}$ is independent from item category, 
 $\boldsymbol{h}_{u,i}^{\perp C}$  should \textit{fool} the discriminator by maximizing the classification loss:
  \begin{equation}
  \max \mathcal{L}_D^{\perp C} (u,i) =  \mathcal{L}_{CE} \left( D(\boldsymbol{h}_{u,i}^{\perp C}), \tilde{\boldsymbol{t}}_i\right). \nonumber
    \end{equation}
 \end{itemize}
  We leverage a Gradient Reverse Layer (GRL) \citep{ganin2015unsupervised} to implement above requirements due to its simplicity. 
  More specifically,  we insert a Gradient Reverse Layer  between 
  $\boldsymbol{h}_{u,i}^{\perp C}$  and the discriminator, as shown in Figure~\ref{fig:dcrs}. 
  During back propagation, the gradients for minimizing the discriminator loss $\frac{\partial \mathcal{L}_D^{\perp C} (u,i)}{\partial\boldsymbol{h}_{u,i}^{\perp C}}$ flow backward through the discriminator. After the GRL, the gradients will be reversed, i.e., becoming $-\frac{\partial \mathcal{L}_D^{\perp C} (u,i)}{\partial\boldsymbol{h}_{u,i}^{\perp C}}$.
  Thus, we perform gradient descent on parameters of the discriminator for accurately predicting item $i$'s category, while performing gradient ascent on $\boldsymbol{h}_{u,i}^{\perp C}$, so that $\boldsymbol{h}_{u,i}^{\perp C}$  cannot predict item $i$'s category.

\noindent
\textbf{Learning category-independent representation.} 
$\boldsymbol{h}_{u,i}^{\perp C}$ should be optimized under two objectives: 
(1) it can accurately estimate the first term $P(Y_{u,i}^{\perp C} =1 | u,i)$ in  Eq.(\ref{equ:P_distenglemnt}) by:
\begin{equation}
  \hat{p}_{u,i}^{\perp C} = P \left(Y_{u,i}^{\perp C}=1 |u,i\right)  = \sigma \left( \boldsymbol{W_1}^{\top} \boldsymbol{h}_{u,i}^{\perp C} \right);
  \label{equ:pui-perpc}
\end{equation}
and (2) it needs to be independent from item categories. 
Thus we minimize the following loss for its learning:
\begin{equation}
    \mathcal{L}_{rec} \left(\hat{p}_{u,i}^{\perp C}, y_{u,i}\right) - \lambda \mathcal{L}_D^{\perp C} (u,i)
  \label{equ:loss_cat_independent}
 \end{equation}
where the two terms optimize two distinct objectives respectively, and
$\lambda$ is a hyper-parameter that controls the strength of category-independent constraint on $\boldsymbol{h}_{u,i}^{\perp C}$ .

\noindent
\textbf{Learning category-dependent representation.}
While user $u$'s preference on item $i$'s category is unobservable, 
$P(Y_{u,i}^C=1|u, t_{i})$ can be estimated by fixing the learnt category-independent representation $\boldsymbol{h}_{u,i}^{\perp C}$  and estimating the overall probability that user $u$ gives  positive feedback to item $i$:  
 \begin{equation}
     \hat{p}_{u,i} = P(Y_{u,i}=1 | u,i)= \sigma \left( \boldsymbol{W_2}^\top \left[ \begin{matrix}
                                  {\rm stop\_gradient}(\boldsymbol{h}_{u,i}^{\perp C}) \\
                                  \boldsymbol{h}_{u,i}^C
                                  \end{matrix} \right]
                                   \right), \boldsymbol{W_2}  \in R^{2d\times 1} 
    \label{equ:pred}
 \end{equation}
where  stop\_gradient$\left( \boldsymbol{h}_{u,i}^{\perp C} \right)$ implies that $\boldsymbol{h}_{u,i}^{\perp C}$ will not be updated by this prediction.
 In other words, given the learnt user $u$'s preference over category-independent features of item $i$, only user $u$'s preference over item $i$'s category is optimized to accurately predict the overall feedback of user $u$ to item $i$, by minimizing the loss:
  \begin{equation}
     \mathcal{L}_{rec} \left(\hat{p}_{u,i}, y_{u,i} \right) + \lambda   \mathcal{L}_D^{C}(u,i),
     \label{eq:loss_cat_dependent}
 \end{equation}
where the second loss forces $\boldsymbol{h}^{C}_{u,i}$ to predict item $i$'s category accurately with $\lambda$ representing the strength of the constraint.
 
 Overall, combining Eq.(\ref{equ:loss_cat_independent}) and Eq.(\ref{eq:loss_cat_dependent}),  given a user behavior dataset $\boldsymbol{\mathcal{X}}$, DCRS learns a disentangled recommendation model as in Eq.(\ref{equ:P_distenglemnt}) by minimizing the following loss:
\begin{align*}
       \mathcal{L}(\boldsymbol{\mathcal{X}}, \boldsymbol{\theta}, \boldsymbol{W_1}, \boldsymbol{W_2}) = &  \frac{1}{|\boldsymbol{X}|}  \sum_{(u, i, y_{u,i})  \in \boldsymbol{\mathcal{X}}}  \mathcal{L}_{rec} \left(\hat{p}_{u,i}, y_{u,i} \right) \nonumber \\ & +\mathcal{L}_{rec} \left(\hat{p}_{u,i}^{\perp C}, y_{u,i}\right) -\lambda \mathcal{L}_D^{\perp C} (u,i) + \lambda  \mathcal{L}_D^{C}(u,i).
\end{align*}

\noindent
\textbf{Inference.}
At the inference stage, we leverage $\hat{p}_{u,i}$ in Eq.(\ref{equ:pred}) as the predicted preference of user $u$ over item $i$ to rank items.
%according to the predicted score. 
We adopt  $\hat{p}_{u,i}$ in  Eq.(\ref{equ:pred}) since it considers both the category dependent and independent preference of the user, while $\hat{p}_{u,i}^{\perp C}$ in  Eq.(\ref{equ:pui-perpc}) only captures the user's preference over category-independent features.

\section{Offline Experiments}
\label{sec:experiments}
In this section, we conduct experiments on several public offline datasets to demonstrate the effectiveness of DCRS.
We mainly investigate from two perspectives:
\begin{itemize}
    \item How does the proposed DCRS perform in terms of recommendation accuracy and diversity?
    \item Can the disentangled category-independent representation accurately distinguish a user's preference within item categories?
\end{itemize}
A case study is also conducted to illustrate the effectiveness of the proposed DCRS more explicitly.
% In this section, we conduct extensive experiments to evaluate performance of DCRS in terms of recommendation accuracy and diversity.
% We also dive deeper to see: (1) whether disentangled category-independent representation can distinguish a user's preference within item category more accurately; and (2) whether disentangled representations can contribute to better generalization in predicting a user's preference on new category.
% A case study is also conducted to illustrate the effectiveness of proposed
% DCRS more explicitly.

\subsection{Experimental Settings}
\label{sub_sec: experimental_setting}
\noindent
\textbf{Dataset.} We use three widely-used datasets under different recommendation scenarios for evaluation.
\begin{itemize}
    \item \textbf{ML-1M\footnote{\url{https://grouplens.org/datasets/movielens/1m/}}:} This dataset contains 1 million ratings from 6040 users on 3883 movies from the online movie recommendation service MovieLens. It also contains rich user features (e.g., age, gender, etc.) and movie features (e.g., titles).
    We encode user and movie features following previous work \citep{zhou2018deep,wang2021deconfounded}. 
    We take $y_{u,i}=1$, if user $u$ gives item $i$ a rating greater than 3, otherwise $y_{u,i}=0$.
    
    \item \textbf{ML-10M\footnote{\url{https://grouplens.org/datasets/movielens/10m/}}:} This dataset is also from MovieLens.
    It contains 10 million ratings from 69878 users on 10680 movies.
    Similarly, we take $y_{u,i}=1$, if user $u$ gives item $i$ a rating greater than 3, otherwise $y_{u,i}=0$.
    
    \item \textbf{Amazon-Books\footnote{\url{https://jmcauley.ucsd.edu/data/amazon/}}:} This dataset contains reviews and metadata of books from Amazon.
    To ensure data quality, we only keep categories that link to more than 20 books with 141 categories, and 
    adopt the 20-core settings \citep{wang2021deconfounded}, i.e., discarding users and books with less than 20 interactions. 
    To make the number of positive and negative samples balanced, we take $y_{u,i}=1$, if user $u$ gives item $i$ a rating greater than 4, otherwise $y_{u,i}=0$.
\end{itemize}
The statistics of the three datasets are summarized in Table~\ref{table:data}.

\begin{table}
\centering
\caption{Statistics of Three Datasets}
\label{table:data}
\begin{tabular}{lllll}
\hline \hline
Dataset  & \#Users & \#Items & \#Interactions & \#Group \\
\hline 
ML-1M & 6040 & 3883 & 1000209 & 18 \\
ML-10M & 69878 & 10680 & 10000047 & 19 \\
Amazon-Books &  22929 & 33130 & 1178117 & 141 \\
\hline
\hline
\end{tabular}
\end{table}
On each dataset, we also randomly sampled  items that the user did not interact with as  negative instances.
We then sorted the user-item interactions by timestamps,  and split them into training, validation, and testing
datasets with the ratio of 80\%, 10\%, and 10\%.

\noindent
\textbf{Baselines.}
The proposed DCRS is a general  and model-agnostic framework to disentangle category dependent and independent representations for accurate and diverse recommendations.
In this paper, we instantiated it with Neural Factorization Machine (NFM)~\citep{he2017neural}, one representative recommendation model that has been widely used.
NFM was also taken as the backbone model in several closely related work for diversified recommendations \citep{grgic2016case,wang2021deconfounded}.
We compared DCRS with the following algorithms that have different focuses on recommendation diversity and accuracy.
\begin{itemize}
    \item \textbf{NFM~\citep{he2017neural}}: The state-of-the-art recommendation model serving as the backbone model of DCRS.
    \item \textbf{Unawareness~\citep{grgic2016case}}: It also takes NFM as the backbone model and tries to improve diversity by directly removing  categorical features of items from model input. 
    \item \textbf{IPS~\citep{saito2020unbiased}}: It is a state-of-the-art technique of improving diversity by boosting item categories that a user interacted with less often, while suppressing the dominant categories in the user's interaction history.
    Specifically, it takes the category distribution in a user's historical interactions as propensity scores to reweigh items of this category during training.
    Propensity clipping \citep{saito2020unbiased} is also employed to reduce the variance with clipping threshold searched in \{0.001, 0.005, 0.01, 0.05, 0.1\}.
    \item \textbf{MMR~\citep{carbonell1998use}}: One of the state-of-art post-processing methods for diversified recommendations.  It re-ranks the recommended items generated by NFM by a greedy strategy to reduce redundancy.
    \item \textbf{DPP~\citep{chen2018fast}}: An effective post-processing method for diversified recommendations.
    It selects a diverse set of items from  the recommended items generated by NFM by balancing the relevance of items and their similarities. 
    \item \textbf{PD\_GAN~\citep{wu2019pd}}: A recent work that leverages the generative adversarial networks (GAN) framework to generate diverse and relevant recommendations.
    Its discriminator aims to distinguish  
    the generated diverse set of items by its generator from the ground-truth
    sets randomly sampled from the observed data of the user.
    \item \textbf{DGCN~\citep{zheng2021dgcn}}: 
    A recent work that leverages rebalanced neighbor discovering, category-boosted negative sampling and adversarial learning on top of Graph Convolutional Networks (GCN) for diversified recommendations.
    \item \textbf{DecRS~\citep{wang2021deconfounded}}: A recent work for alleviating the bias that previous recommendation models over-recommend items of the dominant categories in a user's interaction history from a causal view. It aims at improving both recommendation accuracy and diversity.
    % It started from a causal view and identified that the main reason of above bias issue lies in the confounder effect of imbalanced item distribution on user representation and prediction score.
    % DecRS leverages backdoor adjustment to eliminate the impact of the identified confounder, with NFM as its backbone model.
    \item \textbf{DCRS\_CI}: A variant of DCRS that leverages $\hat{p}_{u,i}^{\perp C}$ in Eq.(\ref{equ:pui-perpc}) for item ranking without considering the user's preference over categories.
    Its comparison with DCRS\_CI can reveal the importance of modeling users' categorical preference.
    %in diversified recommendations. 

\end{itemize}

\noindent
\textbf{Implementation Details.}
Following previous work~\citep{wang2021deconfounded, he2017neural},
we set the embedding size of user/item features to 64 (i.e., $d=64$), and used AdaGrad~\citep{duchi2011adaptive} for optimization.
We used grid search to select the hyperparameters based on the model's performance on validation dataset: 
the learning rate was searched in \{0.005, 0.01, 0.05\};
the normalization coefficient was searched in \{0, 0.1, 0.2\};
the dropout ratio was searched in \{0.2, 0.3, ..., 0.5\};
$\lambda$ for controlling strength of category independent and dependent constraints was searched in \{0.01, 0.05, 0.1, 0.5, 1\}.
For baseline algorithms, when evaluating on the  dataset the algorithms were also evaluated in their original papers, we adopted the recommended hyperparameters from the original paper; otherwise we performed a similar grid search as above with the search range following the original paper.
\subsection{Performance on Recommendation Accuracy \& Diversity}
\label{subsec:overall_performance}
We first evaluate all algorithms in terms of recommendation accuracy and diversity.

\begin{table*}[t]
\centering 
\resizebox{\textwidth}{!}{
\begin{tabular}{||c|c| l l r| l l l l| l l l l||}
  \hline
  \hline  Dataset & Method & AUC & UAUC & RelaImpr & R@10 & NDCG@10 & CE@10 & CC@10 & R@20 &  NDCG@20 & CE@20 & CC@20  \\
  \hline
 \multirow{9}{*}{{\bf ML\_1M}}
   & NFM & 0.8461 & 0.8224 & 0.00\% & 0.0522 & 0.0572 & 1.8056 & 0.4741 & 0.0908 & 0.0681 & 1.9764  & 0.6185\\
  & UnAwareness & $0.8414^-$ & $0.8134^-$ & -2.79\% & $0.0512^-$ & $0.0568^-$ & $1.8919^+$ & $0.4998^+$ & $0.0880^-$ & $0.0669^-$ & $2.0513^+$ & $0.6419^+$\\
  & IPS & $0.8446^-$ & $0.8210^-$ & -0.43\% & $0.0513^-$ & $0.0572$ & $1.7929^-$ & $0.4713^-$ & $0.0890^-$ & $0.0681$ & $1.9759^-$ & $0.6225^+$\\
  & MMR & $\quad -$ & $0.8194^-$ & -0.93\% & $0.0501^-$ & $0.0545^-$ & $2.1279^+$ & $0.5886^+$ & $0.0902^-$ & $0.0670^-$ & $2.2224^+$ & $0.7244^+$\\
   & DPP & $\quad -$ & $0.6021^-$ & -68.3\% & $0.0454^-$ & $0.0518^-$ & $2.4119^+$ & $0.7315^+$ & $0.0770^-$ & $0.0601^-$ & $2.5974^+$ & $0.9586^+$\\
   & PD\_GAN & $\quad -$ & $\quad -$ & $\quad -$ & $0.0326^-$ & $0.0347^-$ & $2.5495^+$ & $0.8347^+$ & $0.0503^-$ & $0.0386^-$ & $2.6650^+$ & $0.9393^+$\\
   & DGCN & $0.7949^-$ & $ 0.7759^-$ & -14.4\% & $0.0365^-$ & $0.0402^-$ & $1.9133^+$ & $0.5088^+$ & $0.0640^-$ & $0.0482^-$ & $2.0748^+$ & $0.6466^+$\\
   & DecRS & $0.8462^+$ & $ 0.8202^-$ & -0.07\% & $0.0537^+$ & $0.0588^+$ & $1.8560^+$ & $0.4876^+$ & $0.0919^+$ & $0.0694^+$ & $2.0378^+$ & $0.6365^+$\\
   \cline{2-13} 
   & DCRS\_CI & $0.8332^-$ & $ 0.8096^-$ & -3.90\% & $0.0530^+$ & $0.0581^+$ & $1.7606^-$ & $0.4468^-$ & $0.0936^+$ & $0.0699^+$ & $1.9108^-$ & $0.5766^-$\\
   & DCRS & $\boldsymbol{0.8483^+}$ & $ \boldsymbol{0.8237^+}$ & \textbf{0.40\%} & $\boldsymbol{0.0551^+}$ & $\boldsymbol{0.0602^+}$ & $\boldsymbol{1.8877^+}$ & $\boldsymbol{0.4909^+}$ & $\boldsymbol{0.0960^+}$ & $\boldsymbol{0.0722^+}$ & $\boldsymbol{2.0620^+}$ & $\boldsymbol{0.6368^+}$\\
   \hline \hline
    \multirow{9}{*}{{\bf ML\_10M}}
   & NFM & 0.8346 & 0.8193 & 0.00\% & 0.0474 & 0.0448 & 1.9351 & 0.5127 & 0.0797 & 0.0547 & 2.0877  & 0.6504\\
  & UnAwareness & $0.8274^-$ & $0.8078^-$ & -3.60\% & $0.0394^-$ & $0.0363^-$ & $2.0308^+$ & $0.5410^+$ & $0.0659^-$ & $0.0446^-$ & $2.2036^+$ & $0.6891^+$\\
  & IPS & $0.8378^+$ & $0.8218^+$ & 0.78\% & $0.0469^-$ & $0.0441^-$ & $1.9280^-$ & $0.5070^-$ & $0.0783^-$ & $0.0538^-$ & $2.0913^+$ & $0.6491^-$\\
  & MMR & $\quad -$ & $0.8084^-$ & -3.41\% & $0.0436^-$ & $0.0418^-$ & $2.2941^+$ & $0.6629^+$ & $0.0762^-$ & $0.0521^-$ & $2.3451^+$ & $0.7639^+$\\
   & DPP & $\quad -$ & $0.6459^-$ & -54.3\% & $0.0390^-$ & $0.0392^-$ & $2.5014^+$ & $0.7740^+$ & $0.0629^-$ & $0.0459^-$ & $2.6248+$ & $0.9376^+$\\
   & PD\_GAN & $\quad -$ & $\quad -$ & $\quad -$ & $0.0108^-$ & $0.0119^-$ & $2.3134^+$ & $0.7164^+$ & $0.0176^-$ & $0.0136^-$ & $2.4446^+$ & $0.8606^+$\\
   & DGCN & $0.8069^-$ & $ 0.8081^-$ & -3.51\% & $0.0425^-$ & $0.0380^-$ & $2.0459^+$ & $0.5530^+$ & $0.0740^-$ & $0.0482^-$ & $2.1925^+$ & $0.6934^+$\\
   & DecRS & $0.8417^+$ & $ 0.8261^+$ & 2.12\% & $0.0477^+$ & $0.0445^-$ & $1.9401^+$ & $0.5048^-$ & $0.0814^+$ & $0.0551^+$ & $2.1181^+$ & $0.6480^-$\\
   \cline{2-13} 
   & DCRS\_CI & $0.8357^+$ & $ 0.8197^+$ & 0.12\% & $0.0478^+$ & $0.0448$ & $1.9810^+$ & $0.5269^+$ & $0.0813^+$ & $0.0554^+$ & $2.1322^+$ & $0.6635^+$\\
   & DCRS & $\boldsymbol{0.8447^+}$ & $ \boldsymbol{0.8301^+}$ & \textbf{3.38\%} & $\boldsymbol{0.0499^+}$ & $\boldsymbol{0.0465^+}$ & $\boldsymbol{2.0050^+}$ & $\boldsymbol{0.5327^+}$ & $\boldsymbol{0.0838^+}$ & $\boldsymbol{0.0572^+}$ & $\boldsymbol{2.1655^+}$ & $\boldsymbol{0.6733^+}$\\
   \hline \hline 
     \multirow{9}{*}{{\bf Amazon-Books}}
   & NFM &  0.6667 & 0.6289 & 0.00\% & 0.0076 & 0.0052 & 1.6722 & 0.0495 & 0.0118 & 0.0066 & 1.9551  & 0.0740\\
  & UnAwareness & $0.6267^-$ & $0.5687^-$ & -46.7\% & $0.0064^-$ & $0.0043^-$ & $1.6660^-$ & $0.0524^+$ & $0.0097^-$ & $0.0054^-$ & $1.8762^-$ & $0.0721^-$\\
  & IPS & $0.6650^-$ & $0.6269^-$ & -1.55\% & $0.0078^+$ & $0.0053^+$ & $1.5969^-$ & $0.0453^-$ & $0.0115^-$ & $0.0066$ & $1.9148^-$ & $0.0704^-$\\
  & MMR & $\quad -$ & $0.6096^-$ & -15.0\% & $0.0067^-$ & $0.0045^-$ & $2.2899^+$ & $0.0864^+$ & $0.0109^-$ & $0.0060^-$ & $2.5119^+$ & $0.1278^+$\\
   & DPP & $\quad -$ & $0.5300^-$ & -76.7\% & $0.0054^-$ & $0.0040^-$ & $2.5184^+$ & $0.1005^+$ & $0.0081^-$ & $0.0049^-$ & $2.8741+$ & $0.1645^+$\\
   & PD\_GAN & $\quad -$ & $\quad -$ & $\quad -$ & $0.0004^-$ & $0.0003^-$ & $2.3179^+$ & $0.0920^+$ & $0.0016^-$ & $0.0007^-$ & $2.7648^+$ & $0.1545^+$\\
   & DGCN & $0.6747^+$ & $ 0.6404^+$ & 8.92\% & $0.0071^-$ & $0.0044^-$ & $2.0003^+$ & $0.0698^+$ & $0.0122^+$ & $0.0061^-$ & $2.2842^+$ & $0.1074^+$\\
   & DecRS & $0.6964^+$ & $ 0.6558^+$ & 20.8\% & $0.0074^-$ & $0.0051^-$ & $1.8207^+$ & $0.0601^+$ & $0.0111^-$ & $0.0063^-$ & $2.0973^+$ & $0.0918^+$\\
   \cline{2-13} 
   & DCRS\_CI & $0.6893^+$ & $ 0.6546^+$ & 19.9\% & $0.0057^-$ & $0.0036^-$ & $2.0172^+$ & $0.0704^+$ & $0.0095^-$ & $0.0049^-$ & $2.2799^+$ & $0.1068^+$\\
   & DCRS & $\boldsymbol{0.6974^+}$ & $ \boldsymbol{0.6573^+}$ & \textbf{22.0\%} & $\boldsymbol{0.0079^+}$ & $\boldsymbol{0.0052}$ & $\boldsymbol{1.8639^+}$ & $\boldsymbol{0.0622^+}$ & $\boldsymbol{0.0123^+}$ & $\boldsymbol{0.0067^+}$ & $\boldsymbol{2.1415^+}$ & $\boldsymbol{0.0953^+}$\\
  \hline \hline 
\end{tabular}
}
\caption{Experimental results regarding to recommendation accuracy and diversity. Improved (or dropped) performance over the base NFM model under the same setting is marked as  $+$ (or $-$). }
\label{table:exp1}
\resizebox{\textwidth}{!}{
\begin{tabular}{||c|c| r r r r|r r r r|r r r r||}
  \hline
  & & \multicolumn{4}{|c|}{{\bf ML-1M }} & \multicolumn{4}{|c|}{ {\bf  ML-10M} } & \multicolumn{4}{|c|}{\bf Amazon-Books} \\
  \hline \hline
  Category & Method & AUC & UAUC  & R@20 & NDGG@20  & AUC & UAUC  & R@20 & NDGG@20   & AUC & UAUC  & R@20 & NDGG@20  \\
  \hline 
  \multirow{3}{*}{1$^\text{st}$ ranked cat} 
   & NFM & 0.8547 & 0.8180 & 0.3034 & 0.1678  & 0.8498 & 0.8229   & 0.2814 & 0.1453 & 0.6474 & 0.5976  & 0.0679& 0.0343  \\
   & DecRS & 0.8563 & 0.8135 & 0.3079 & 0.1718  &  0.8545 & 0.8273 & 0.3031 & 0.1572 & 0.6724 & 0.6113   & 0.0698 & 0.0331\\
   & DCRS\_CI & \textbf{0.8606} & \textbf{0.8241} & \textbf{0.3230} &  \textbf{0.1783}& \textbf{0.8608} & \textbf{0.8340} & \textbf{0.3210}& \textbf{0.1659}  & \textbf{0.6730} & \textbf{0.6178}  & \textbf{0.0730} & \textbf{0.0345} \\
  \hline \hline 
  \multirow{3}{*}{2$^\text{nd}$ ranked cat} 
  & NFM & 0.8403 & 0.8009  & 0.3820 & 0.1962 & 0.8372 & 0.8078  & 0.3434 & 0.1655 & 0.6637 & 0.5489   & 0.0536 & 0.0302\\
  & DecRS &0.8407 & 0.8014 & 0.3817 & 0.1982&  0.8407 & 0.8088  & 0.3558 & 0.1718 & 0.6978 & 0.5661   & 0.0576 & 0.0310 \\
  & DCRS\_CI & \textbf{0.8449} & \textbf{0.8056}  & \textbf{0.3960} & \textbf{0.2078} & \textbf{0.8485} & \textbf{0.8170} & \textbf{0.3861}& \textbf{0.1875} & \textbf{0.7413} & \textbf{0.5709}  & \textbf{0.0595} & \textbf{0.0312} \\
  \hline \hline
  \multirow{3}{*}{3$^\text{rd}$ ranked cat} 
  & NFM & 0.8344 & 0.8046 & 0.6665 & 0.3350 & 0.8172 & 0.7926& 0.4156 & 0.1969 & 0.6931 & 0.5910  & 0.0554 & 0.0241 \\
  & DecRS & 0.8381 & \textbf{0.8062}  & 0.6743 & 0.3423 &  0.8231 & 0.7968  & 0.4458 & 0.2137& 0.7044 & 0.5920 & 0.0548 & 0.0233\\
  & DCRS\_CI & \textbf{0.8419} & 0.8055 & \textbf{0.6873} & \textbf{0.3463} & \textbf{0.8264} & \textbf{0.8014} & \textbf{0.4794} & \textbf{0.2291} & \textbf{0.7176} & \textbf{0.6162}   & \textbf{0.0591} & \textbf{0.0245} \\
  \hline
\end{tabular}}
\caption{Recommendation accuracy of disentangled category-independent representation on category-specifc testing data.}
\label{table:exp2}
\vspace{-10mm}
\end{table*}

\noindent
\textbf{Evaluation Metrics.}
We evaluate the accuracy of a recommendation model from two perspectives:
(1) Whether the model can rank positively interacted items of a user before those negatively interacted ones accurately in the testing dataset;
(2) Whether the model can accurately retrieve those positively interacted items in the testing dataset from the item pool, which includes all items that the user did not interact with in the training dataset.
For MMR and DPP, because they only re-rank the recommended items generated by NFM, a specifically created item pool that contains top-200 items of NFM is used.
We adopted AUC~\citep{fawcett2006introduction} and UAUC~\citep{zhou2018deep} as metrics to evaluate the first perspective.
Basically, UAUC is a micro-average version of AUC, measuring the goodness of intra-user recommendation by averaging AUC over users.
Besides, we followed previous work~\citep{yan2014coupled, zhou2018deep} to use the RelaImpr metric to measure the relative improvement over the base NFM model on UAUC.
For a random guesser, the value of AUC is 0.5, and thus RelaImpr is defined as:
\begin{equation}
    {\rm RelaImpr} = \left( \frac{{\rm UAUC(mesured \ model)} -0.5}{{\rm UAUC(base \ NFM \ model)}-0.5} -1 \right ) \times 100\% \nonumber.
\end{equation}
To evaluate the second perspective of recommendation accuracy, we adopted  Recall@K and NDCG@K for the purpose.
Regarding recommendation diversity, we used two widely-adopted metrics:   (1) Category coverage (CC@K), which is the ratio between number of categories covered by top-K recommendations and the total number of categories in dataset; (2) Category entropy (CE@K), which is the entropy of category distribution in top-K recommendations. 
Higher CC@K and CE@K  suggest more diverse top-K recommendations.

Table~\ref{table:exp1} shows the experiment results of all algorithms. 
We cannot report AUC, UAUC and RelaImpr for PD\_GAN, since it directly recommends a set of items.
For MMR and DPP, we can only report  UAUC and RelaImpr since it is hard to find an appropriate way to merge the re-ranked list of different users to calculate AUC.
Based on the results, we can observe that:
\begin{itemize}
\item Although Unawareness, MMR, DPP, PD\_GAN and DGCN promoted more diverse recommendations with higher CE@K and CC@K, their recommendation accuracy dropped a lot, indicating their failure to handle accuracy-diversity dilemma. 
\item IPS did not consistently outperform the base NFM model in recommendation diversity or accuracy, due to the inaccurate estimation and high variance of propensity scores.
\item At most time, especially on ML\_1M and ML\_10M100K dataset, DecRS improved both recommendation accuracy and diversity, since it could avoid many less-relevant or low-quality items from the dominant categories being recommended. 
However, its improvement was not larger than our proposed DCRS.
\item Our proposed DCRS effectively improved both recommendation accuracy and diversity on all three datasets compared to the base NFM model.
One can observe on all datasets, DCRS achieved the highest recommendation accuracy in all metrics, and generated more diverse recommendations than the base NFM model with higher CC@K and CE@K. 
This implies that DCRS tends to generate diverse recommendations the users will prefer, rather than solely pursuing diversity regardless of recommendation accuracy.
%DCRS achieves this by  diversifying recommendations only within item categories that a user may prefer through disentangling the user's category dependent and independent preference.
Moreover, compared to DCRS, the recommendation accuracy of DCRS\_CI dropped on all three datasets, confirming the importance of modeling users' categorical preference. 
\end{itemize}

%\subsection{Performance of Disentangled Category-Independent Representation}
\subsection{Predicting Users' In-Category Preferences}
\label{sec:cat_independent_eval}
We dive deeper to investigate why DCRS can make accurate and diversified recommendations.
Based on our design, the disentanglement shields the users' preference on item categories from their preference on items within the category when learning the user-item representations. As a result, the user-item representations learnt by DCRS 
%Intuitively, compared to user-item representations obtained without disentanglement, the disentangled category-independent representation is not distorted by the user's preference on item categories, thus it 
should better predict a user's interest within item category, compared to those did not consider this aspect.
Thus we inspect whether the disentangled category-independent representation (i.e.,  $\{\boldsymbol{h}_{u,i}^{\perp C}\}$) can distinguish less relevant (or low-quality) items from relevant (or high-quality) items more accurately within a given category of items.

We split the testing dataset according to item categories, and evaluated all algorithms on each category-specific testing dataset separately.
On all three of our evaluation datasets, an item may relate to multiple categories. For example, the movie ``Toy Story (1995)'' in ML-1M dataset is related to three categories: ``Animation'', ``Children's'', and ``Comedy''.
Here, we split the testing dataset according to each unique combination of related categories.
Then given one unique combination of categories, we traversed the testing dataset and only kept user-item interactions where the interacted item is associated with the same category combination.
To ensure the reliability of the evaluation results, on each dataset, we only evaluated the algorithms on the top-3 most popular categories.

% For inspecting the effectiveness of disentangled category-independent representation (i.e.,  $\{\boldsymbol{h}_{u,i,\perp c}\}$),
In this experimental setting, we only need to evaluate DCRS\_CI, as all items are from the same category.
Table~\ref{table:exp2} demonstrates the experiment results.
Due to space limit, we only report results on AUC, UAUC, Recall@20 and NDCG@20, and omit baselines that perform worse than the base NFM model.
From Table~\ref{table:exp2}, we can observe that 
both DecRS and DCRS\_CI performed better than the base NFM model, as aligned with the results in Section~\ref{subsec:overall_performance}.
Moreover, DCRS\_CI achieved the best performance  most time, implying that
disentangled representations contribute to more accurate preference modeling within categories.

\subsection{Case Study}
\begin{figure}[t]
    \centering
    \includegraphics[width=0.8\textwidth]{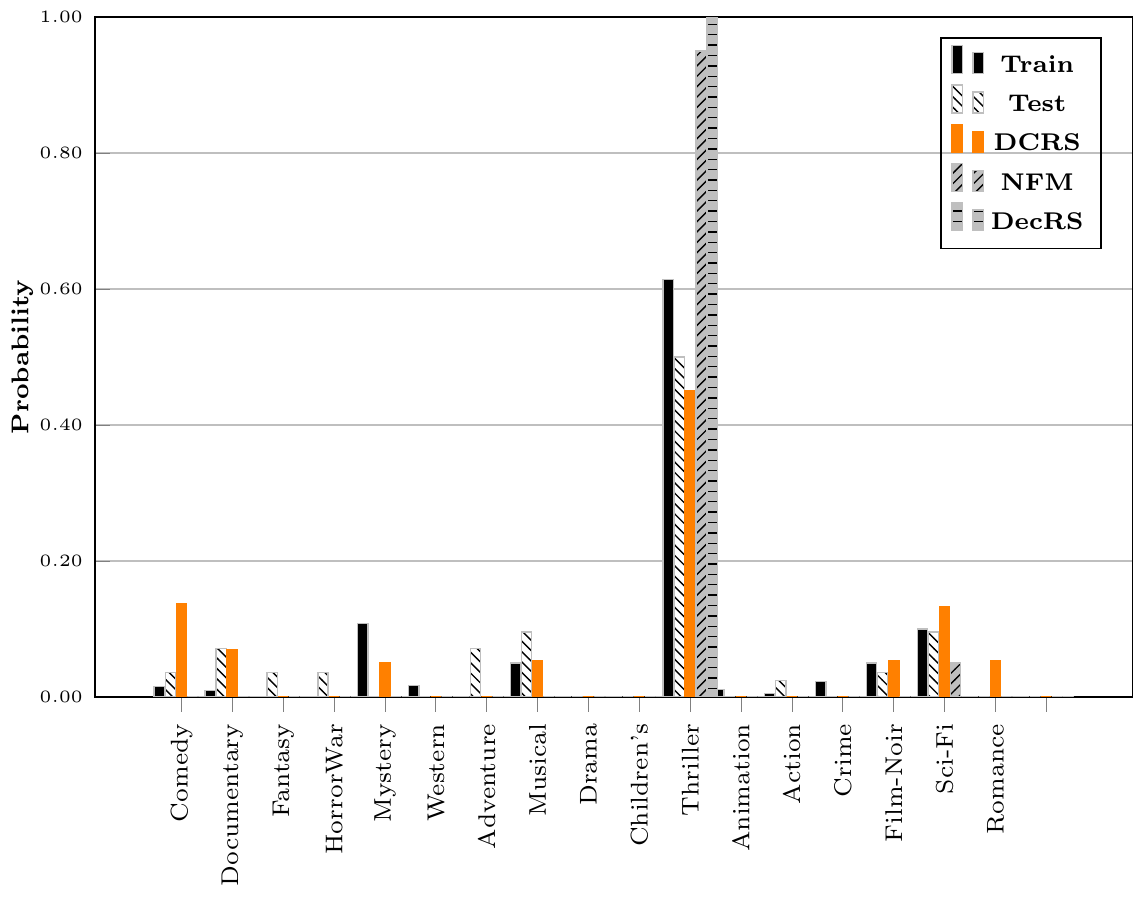}
    \caption{Categorical distributions of training data, testing data and  top-10 recommended items of a sampled user.}
    \label{fig:case_study}
    \vspace{-6mm}
\end{figure}

We also use a case study to qualitatively illustrate the behavior of the proposed DCRS model.
%We select the testing users from ML-1M dataset where we observe the largest discrepancy between the distribution of categories in the top-10 recommendations generated by NFM and DCRS. 
Figure~\ref{fig:case_study} shows the distribution of categories in the interacted items in training and testing data of a user from ML-1M dataset, as well as the top-10 recommended items generated by NFM, DecRS and DCRS.
One can observe from Figure~\ref{fig:case_study} that:
the top-10 recommendations of NFM and DecRS model mainly fell in the ``Thriller'' category, which is the most popular in the training data of this user.
Our proposed DCRS could capture the user's preference over categories more thoroughly. 
As shown in Figure~\ref{fig:case_study}, the recommended items from DCRS did not simply concentrate to the dominant category ``Thriller'' as in other recommendation algorithms, but they successfully covered six out of ten categories that have a non-zero support in the user's testing data.
    % And we want to remark that the user did not interact with items in the left three categories may just due to no recommendations in these categories. 
Moreover, DCRS could also identify the user's preference on categories that the user seldom interacted with before, for example, the category of ``Documentary''. This explains its improved recommendation diversity without losing recommendation accuracy.
    % This explicitly demonstrate the effectiveness of DCRS:
    % it can capture the user's diverse preference in historical interactions more thoroughly, and also rank items in the same category more accurately, leading to 

\section{Online Deployment and A/B Test}
To further verify the effectiveness of DCRS, we deploy it on the recommendation channel of Toutiao app, one of the largest news recommendation platforms in China, for online A/B test. 

More specifically, we implemented DCRS based on the main candidate generator of Toutiao.
Here, the main candidate generator
%\hnote{how about ``candidate retrieval module''}
is one of many candidate generators that produce recommendation candidates, which are later scored and ranked by a separate ranking model before presenting to users~\citep{chen2019top}. 
But the recommendation candidates produced by the main candidate generator account for the largest proportion of the recommendations shown to users. 
We then replaced the main candidate generator by DCRS in the experimental group, and used the prior main candidate generator in the control group.
We adopted two key metrics: (1) \textit{Click Through Rate (CTR)}; (2) \textit{StayTime}, to measure users' satisfaction with the resulting recommendations.
To accurately evaluate recommendation diversity, 
%it would be pretty noisy if all displayed items are considered, because some items, especially those displayed less often, are more likely recommended by some special strategies. For noise reduction, 
we only targeted items with more than 1000 impressions, because for those that appear less frequently could be introduced by some special strategies rather than the compared methods.
We then calculated four metrics: (1) E\_CN: number of distinct categories of targeted items shown to a user; (2) E\_CE: entropy of category distribution of targeted items shown to a user ; (3) R\_CN: number of distinct categories of targeted items read by a user ; (4) R\_CE: entropy of category distribution of targeted items read by a user. 
The A/B test was conducted for seventeen consecutive days and the average performance of the above metrics is reported.
We report experimental results in Table~\ref{table:online_exp}.
All reported results are significant with p-value $<0.05$.
We can observe that DCRS achieved higher  \textit{CTR} and \textit{StayTime}, indicating improved users' satisfaction.
Moreover, while the improvements in E\_CN and E\_CE were not that large, DCRS gained  huge improvements in R\_CN and R\_CE, implying DCRS is able to generate \textit{diverse
recommendations the user will prefer.}

\begin{table}
\centering 
\resizebox{0.8\linewidth}{!}{
\begin{tabular}{||c| r r|r r |r r||}
  \hline  & $\Delta${CTR} & $\Delta$\textit{StayTime} & $\Delta$E\_CN & $\Delta$E\_CE & $\Delta$R\_CN & $\Delta$R\_CE  \\
  \hline 
   DCRS & +0.973\% & +0.062\% &  +0.197\% & +0.111\% & +2.372\% & +2.276\%  \\
  \hline \hline
\end{tabular}}
\caption{Results of online A/B test on Toutiao app.}
\label{table:online_exp}
\vspace{-6mm}
\end{table}

\section{Relate work}
DCRS is closely related to two lines of existing work:
%aims to disentangle category dependent and
%independent representation for accurate and diversified recommendations,
(1) addressing accuracy-diversity dilemma in recommendations; and
(2) disentangled user representation learning for general user modeling.

\noindent
\textbf{Addressing accuracy-diversity dilemma in recommendations.}
Besides recommendation accuracy, 
%While improving  recommendation accuracy  has been taken as a primary goal in early recommendation research, 
more and more research suggests other factors of recommendation quality also
contribute to the overall user satisfaction about the system.
Of these factors, recommendation diversity has been shown as a critically important one~\citep{anderson2020algorithmic,wilhelm2018practical}, which however also leads to the so-called
accuracy-diversity dilemma~\citep{wang2021deconfounded,zheng2021dgcn}: higher accuracy often means losing diversity to some extent and vice verse.
One main reason is that previous solutions with accuracy as the primary goal tend to focus on items in the dominant categories in users’ interaction history.
In order to guarantee user satisfaction, three different types of solutions are proposed, namely post-processing, 
learning to rank, and diversified recommendation models.

For the first, and most popular, type of solutions,  a re-ranking or post-processing module is appended to a chosen recommendation model. 
The post-processing module takes  recommended items as input and re-orders them to balance recommendation accuracy and diversity.
Various post-processing algorithms \citep{ziegler2005improving,qin2013promoting,ashkan2015optimal,chen2018fast,kaya2019comparison} are proposed.
For example, Ziegler et. al.~\citep{ziegler2005improving} first applied the Maximal Marginal Relevance (MMR) algorithm,  which was used for topic diversification in search engines,  to minimize redundancy among recommended items. 
%Qin et. al.~\citep{qin2013promoting} proposed an entropy regularizer to capture
%the notion of diversity, and linearly combined it with predictions generated by regular recommendation models for re-ordering items. 
%AshKan et. al.~\citep{ashkan2015optimal} then replaced the linear combination by multiplication to calculated final scores for re-ranking.
%Sha et. al~\citep{sha2016framework} further toke  the coverage of user interest into account when balancing recommndation diversity and accuracy.
 Determinantal Point Process (DPP) has been shown as the most effective one \citep{chen2018fast} of all post-processing algorithms, which scores an entire list of items rather than every item individually for better modeling of item correlations.
However, all these post-processing algorithms are separately constructed from the recommendation models, though their learning highly depends on the performance of the recommendation model.
When the recommendation model fails to provide a diverse item list to start with, or gives pretty-low scores to diverse items,
the effectiveness of the aforementioned post-processing algorithms will largely deteriorate.
 Moreover, as shown in our experiment results in Section~\ref{sec:experiments}, the aforementioned post-processing algorithms usually seriously sacrifice recommendation accuracy.

% Our proposed DCRS aims to improve both recommendation accuracy and diversity of the recommendation model by  disentangling users' preferences.
% It is orthogonal to the line of work on post-processing, but it may contribute to post-processing by providing more diverse top-K recommendations.

Learning To Rank type solutions  \citep{cheng2017learning,wu2018sql,liu2022determinantal} aim to directly recommend a list of items to users, rather than selecting items one by one according to their prediction scores.
However, this line of work often suffers from high time complexity, which limits its application in real world recommendation scenarios.

Recently, several solutions are proposed to directly improve the diversity of recommendation models.
Zheng et. al~\citep{zheng2021dgcn} proposed a diversified recommendation model based on Graph
Convolutional Networks (GCN), with improving recommendation diversity as its only target.
Wu et. al~\citep{wu2019pd} leveraged the GAN framework for diverse recommendations, where a generator tries to recommend diverse sets of items, and a discriminator aims to distinguish the generated recommendations from a set of items randomly sampled from the observed data of
the target user.
%Since this work is specifically designed for GCN-based recommendation models, while we take the NFM model as the backbone in this paper, we did not compare with it in our experiments. We want to remark that the proposed DCRS is a general framework, and one can also incorporate it into the model proposed in \citep{zheng2021dgcn} to further improve its recommendation accuracy and diversity.
The most related work to ours is \citep{wang2021deconfounded}, where the authors studied the problem of lack of diversity in recommendations
%why previous recommendation models overrecommend the most popular category in users' interaction history 
from a casual perspective, and proposed DecRS to alleviate the problem. 
%thby leveraging backdoor
%adjustment to eliminate the impact of the hidden confounder.
Experiments demonstrate the advantage of our proposed DCRS over these solutions in improving recommendation accuracy and diversity.
A recent work  \citep{lin2022feature} also tried to  diversify recommendations in relevant recommendation scenario, where the diversification is conducted regarding multiple item aspects such that relevance and diversity are adaptively balanced among different item aspects.  
However, when only one item aspect is considered, e.g., the item category in this paper, their algorithm degenerates to the MMR algorithm.

\noindent
\textbf{Disentangled user representations.}
%User behavior data in recommender systems are driven by complex interactions
%of many latent factors behind users' decision making process.
Learning disentangled user representations has drawn increasing attention in recent years.
%since it can bring
%enhanced robustness, interpretability, and controllability in various user-centric downstream applications, such as recommendation.
A family of solutions are based on Variational Auto-Encoder (VAE) to force each dimension of learnt representations to focus on different latent factors \citep{ma2019learning,xie2021adversarial,nema2021disentangling}.
However, such a disentanglement is implicit and therefore one cannot associate the disentangled representation with the specific attributes of interest. 
 Zheng et. al~\citep{zheng2021disentangling} proposed DICE to learn representations where user interest and conformity are structurally disentangled via direct supervision from cause-specific data.
However, in our problem, we cannot access users' preferences over item categories explicitly, thus are not able to get any direct supervision about it.
Chen et. al. \citep{chen2022co} proposed to disentangle item representations to address popularity bias, by requiring the two disentangled item representations to be orthogonal.
In our solution, we disentangle a user's preference over an item into  category dependent and independent segments. Both segments relate to the user and thus they do not need to be orthogonal to each other.

\vspace{-4mm}
\section{Conclusion}
In this paper, we propose a new principle that the diversification of recommendations should be performed within a user's preferred categories, such that improved recommendation diversity can be achieved without sacrificing recommendation accuracy. We realize this principle via a general framework, named DCRS,  to disentangle a user's category dependent and independent preference in the learnt representations.
We evaluate DCRS through both offline experiments on three widely-used benchmark datasets for recommendation and online A/B test on Toutiao,  one of the largest news recommendation platforms in China.
We demonstrate DCRS can provide more accurate and diversified recommendations.
Via in-depth analysis and case studies, we find that the benefit of DCRS is introduced by: (1) it can capture a user's diverse preference in historical
interactions more thoroughly; and
(2) it can rank items in the same category more accurately.

In this work, we took a static view of users' preferences over items and item categories. But numerous studies have demonstrated that users' preferences evolve over time \cite{wu2018learning,wu2019dynamic}. It is interesting to study the problem of recommendation diversification in an interactive manner over time. Moreover, currently we only recommend one item a time to a user. It is interesting to study how to generate a list of diverse recommendations, where diversity should be optimized within and across recommendation lists.

% \vskip 0.2in
\bibliography{sample}

\end{document}